\documentclass[letterpaper]{article}

\usepackage{flairs}
\usepackage{times}
\usepackage{helvet}
\usepackage{courier}
\usepackage{amsmath} 
\usepackage{tcolorbox} 
\usepackage{booktabs}

\title{Bridging Expectation Signals: LLM-Based Experiments and a Behavioral Kalman Filter Framework}
\author{
Yu Wang\\
\\
\\
\And Xiangchen Liu\\
Department of Family and Consumer Sciences\\
California State University, Long Beach\\
Long Beach, USA\\
}

\begin{document}
\maketitle

\begin{abstract}
As LLMs increasingly function as economic agents, the specific mechanisms LLMs use to update their belief with heterogeneous signals remain opaque. We design experiments and develop a Behavioral Kalman Filter framework to quantify how LLM-based agents update expectations, acting as households or firm CEOs, update expectations when presented with individual and aggregate signals. The results from experiments and model estimation reveal four consistent patterns: (1) agents' weighting of priors and signals deviates from unity; (2) both household and firm CEO agents place substantially larger weights on individual signals compared to aggregate signals; (3) we identify a significant and negative interaction between concurrent signals, implying that the presence of multiple information sources diminishes the marginal weight assigned to each individual signal; and (4) expectation formation patterns differ significantly between household and firm CEO agents. Finally, we demonstrate that LoRA fine-tuning mitigates, but does not fully eliminate, behavioral biases in LLM expectation formation.
\end{abstract}

\section{Introduction}

As Large Language Models (LLMs) are increasingly deployed as autonomous agents for economic forecasting and policy simulation \cite{park2023generative,horton2023,korinek2023generative,lopezlira2023,xi2023challenges}, understanding their internal expectation formation mechanisms becomes critical. Despite their predictive prowess, the ``black box" nature of how these agents integrate conflicting information streams remains poorly understood \cite{li2023ntire,bommasani2021opportunities}. 

On the economic front, while information economists have studied expectation formation for decades, empirical evidence on how heterogeneous shocks interact remains surprisingly scarce \cite{mankiw2002sticky,coibion2012what}. Despite documentation that various expectations are likely formed jointly (\cite{bruinedebruin2011expectations,carroll2003macroeconomic}), the specific mechanisms between concurrent signals have not been formally quantified. 

This paper bridges these gaps by quantifying LLM-based expectation formation through a combined theoretical and empirical framework. We design a large-scale controlled experiment (720 trials) to isolate the influence of dual heterogeneous signals on future expectations. By estimating these dynamics through a reduced-form Behavioral Kalman Filter, we formally document the weighting mechanisms and cognitive biases inherent in model-generated forecasts.

We evaluate LLM expectation formation within a controlled framework where agents integrate dual information signals under two distinct personas: Household and Firm CEO. These personas are selected because households and firms constitute the fundamental decision-making units in economic theory, making their belief-updating mechanisms a critical first step for agent-based simulation \cite{akerlof2000economics}. Furthermore, economic literature suggests significant heterogeneity in expectation patterns between these groups \cite{coibion2018firms}, allowing us to test how identity priming modulates the model's perceived signal-to-noise ratio.

The core experimental task requires LLMs to generate future growth expectations—specifically, household income growth for the Household persona and real profit growth for the CEO persona. Each agent is presented with two exogenous signals: a \textit{micro-level shock} (reflecting idiosyncratic personal income or firm performance) and a \textit{macro-level shock} (reflecting aggregate national GDP growth). This dual-signal structure is designed to evaluate how models reconcile specific, localized information with broader economic indicators during the belief-updating process \cite{lucas1973expectations}.

We evaluate our framework through a large-scale controlled simulation across three state-of-the-art LLMs (GPT-4o, Gemini 1.5, and DeepSeek-V3). By systematic variation of micro-level and macro-level shocks under Household and CEO personas, we construct a dataset of 720 unique experimental trials.

The experimental results yield three pivotal findings. First, when forecasting future profitability, both household and CEO personas prioritize micro-level signals. Second, the CEO persona assigns significantly greater weight to macro signals than the household persona. Third, while specific mean and variance estimates vary across different LLMs, the underlying behavioral patterns remain remarkably consistent.

Motivated by these patterns, we extend the traditional Kalman Filter \cite{kalman1960new} into a Behavioral Kalman Filter (BKF) framework by incorporating two key mechanisms: a prior discounting factor applied to the previous state, and a subjective covariance term between signals. While standard information and macroeconomics typically assume a zero-covariance Kalman Filter, our extensions are designed to capture (1) the cognitive interactions between concurrent signals, and (2) the heterogeneity of signal processing in expectation formation.

Estimating a full structural model can be complex; therefore, we employ a Bayesian linear regression to estimate our BKF, which is better suited for our small sample size \cite{bishop2006pattern}. Specifically, we regress current expectations on the prior expectation, both signals, and their interaction term. Our results show that the sum of the estimated coefficient for the prior and signals is significantly different from one, confirming that all models exhibit irational when new information arrives.

Consistent with our experimental findings, both the household and CEO personas assign a higher weight to micro signals than to macro signals. Furthermore, the negative coefficient of the interaction term provides empirical evidence for the ``cognitive discount" effect: the presence of concurrent signals creates interference that weakens the model's trust in each individual source. Finally, across personas, we confirm a ``professional sensitivity premium," as the CEO persona places significantly more weight on macro signals than households do.

Finally, we examine whether fine-tuning LLMs can eliminate the behavioral component in our BKF framework. We apply Low-Rank Adaptation (LoRA) to Qwen2.5-7B-Instruct and compare responses from the fine-tuned and base models \cite{hu2022lora}. A rational Kalman Filter is defined by two conditions: (1) the weights on all components sum to one, and (2) the interaction term equals zero \cite{coibion2015information,lorenzoni2009theory,durbin2012time}. Our results show that fine-tuning reduces the volatility of model responses. For the household scenario, the interaction term becomes statistically insignificant, while for the CEO scenario it remains significantly different from zero. Moreover, the estimated coefficients do not sum to one. These findings suggest that LoRA fine-tuning mitigates but does not fully eliminate the irrational component in LLM expectation formation.

\section{Theoretical Framework: Behavioral Kalman Filter}

We start by laying out the theoretical framework of the standard Kalman Filter, then introduce the behavioral extension of it. By focusing on how multiple information signals interact, we challenge the rational expectation formation assumptions prevalent in the Economics literature, specifically the principle of additive independence between orthogonal shocks.

\subsection{Standard Kalman Filter Baseline}
To motivate the Behavioral Kalman Filter, we begin with the classical Kalman filtering framework in information economics \cite{phelps1970microeconomic}. Let $x_t$ denote the latent economic state (e.g., future income growth or profitability). Instead of treating signals in isolation, we define the observation as a vector $\mathbf{s}_t = [s_t^{mic}, s_t^{mac}]^\top$ that provides a noisy measurement of the latent state:

\begin{equation}
\mathbf{s}_t = \mathbf{H} x_t + \boldsymbol{\varepsilon}_t, \qquad \boldsymbol{\varepsilon}_t \sim \mathcal{N}(\mathbf{0}, \mathbf{\Sigma}),
\end{equation}

\noindent where $\mathbf{H} = [1, 1]^\top$ is the observation matrix and $\mathbf{\Sigma} = \text{diag}(\sigma_{mic}^2, \sigma_{mac}^2)$ is the noise covariance matrix. Given a prior estimate $x_{t|t-1}$, the classical Kalman filter updates beliefs according to:

\begin{equation}
x_{t|t} = x_{t|t-1} + \mathbf{K}_t^\top \left( \mathbf{s}_t - \mathbf{H} x_{t|t-1} \right),
\end{equation}

\noindent where $\mathbf{K}_t = [K_{t,mic}, K_{t,mac}]^\top$ is the Kalman gain vector. In the standard formulation, the gains are determined by the relative precision of the signals:

\begin{equation}
\mathbf{K}_t^\top = P_{t|t-1} \mathbf{H}^\top \left( \mathbf{H} P_{t|t-1} \mathbf{H}^\top + \mathbf{\Sigma} \right)^{-1}.
\end{equation}

\noindent in which $P_{t|t-1}$ denotes the prior estimation error covariance, representing the agent's uncertainty before observing the current signals. 

This vector structure imposes two strong restrictions: (i) signals are weighted symmetrically based solely on the statistical noise covariance $\mathbf{R}$, and (ii) the update rule is strictly linear, assuming no interaction between the elements of $\mathbf{s}_t$ beyond their shared mapping to $x_t$. 

\subsection{Behavioral Kalman Filter: Model Definition}

To quantify cognitive biases in expectation formation, we define the \textit{Behavioral Kalman Filter} (BKF). In this framework, the agent observes a vector of information shocks $\mathbf{s}_t$, which follows the dynamic:

\begin{equation}
\mathbf{s}_t = \mathbf{H} x_t + \boldsymbol{\eta}_t, \quad \boldsymbol{\eta}_t \sim \mathcal{N}(\mathbf{0}, \mathbf{\Sigma}_b)
\end{equation}

\noindent where $\mathbf{R}_b$ is the subjective noise covariance matrix that governs how the agent perceives the uncertainty and correlation between micro and macro signals:

\begin{equation}
\mathbf{\Sigma}_b = \begin{bmatrix} 
\sigma_{mic}^2 & \rho \sigma_{mic} \sigma_{mac} \\ 
\rho \sigma_{mic} \sigma_{mac} & \sigma_{mac}^2 
\end{bmatrix}.
\end{equation}

The parameter $\rho$ represents the subjective signal correlation, capturing the cognitive interference where the interpretation of one channel is endogenously "polluted" by the state of the other \cite{sims2003implications}. The agent then updates its belief $x_{t|t}$ according to the following recursive rule:

\begin{equation}
x_{t|t} = \alpha x_{t|t-1} + \mathbf{G}_t^\top (\mathbf{s}_t - \mathbf{H} x_{t|t-1})
\end{equation}

\noindent where $\alpha \in [0, 1]$ is the prior discounting factor, representing the decay of the previous state $x_{t|t-1}$. The behavioral gain vector $\mathbf{G}_t$ is derived from the subjective covariance structure:

\begin{equation}
\mathbf{G}_t^\top = P_{t|t-1} \mathbf{H}^\top \left( \mathbf{H} P_{t|t-1} \mathbf{H}^\top + \mathbf{\Sigma}_b \right)^{-1}.
\end{equation}

In this specification, non-rationalities are formally characterized by two parameters: $\rho \neq 0$, which breaks the independence of signal processing, and $\alpha < 1$, which reflects a behavioral bias toward recent information at the expense of prior anchoring.

\subsection{Empirical Strategy: Bayesian Reduced-Form Regression}

Estimating a full structural model can be complicated. In this paper, we estimate a reduced-form regression which allows us to measure how different types of expectation shocks affect expectation formation. This approach maps our theoretical framework to a linear specification that is easier to identify while maintaining mathematical rigor. 

For each LLM agent, we pool observations across all experimental scenarios and estimate the following specification:

\begin{equation}
\begin{split}
x_{t|t, i} ={} & \beta_{prior} x_{t|t-1, i} + \beta_{mic} s_{t, i}^{mic} + \beta_{mac} s_{t, i}^{mac} \\
& + \beta_{int}s_{t, i}^{mic} \cdot s_{t, i}^{mac} + \varepsilon_{t, i}
\end{split}
\end{equation}

\noindent where $i$ indexes each individual data point across all experimental scenarios; $s_{t, i}^{mic}$ and $s_{t, i}^{mac}$ denote the micro and macro expectation shocks, respectively; and $\varepsilon_{t, i} \sim \mathcal{N}(0, \sigma^2)$ represents the idiosyncratic noise in the model's response.

Based on this regression specification, the coefficients provide a direct mapping to the behavioral parameters of the BKF framework. The coefficient $\beta_{prior}$ identifies the prior discounting factor $\alpha$; specifically, its value indicates the degree to which the agent's prior belief $x_{t|t-1}$ anchors or decays when exposed to new information. Following this, the coefficient $\beta_{mic}$ (and similarly $\beta_{mac}$) quantifies the marginal impact of the micro signal $s_{i}^{mic}$ on expectation formation, conditioned on the presence of the macro signal $s_{i}^{mac}$. Crucially, the interaction coefficient $\beta_{int}$ captures the subjective signal correlation $\rho$, explaining how these two distinct information channels non-linearly interfere with one another during the cognitive integration process.

We employ Bayesian linear regression to estimate this model. This choice is motivated by the recursive Bayesian nature of the Kalman Filter and allows us to derive full posterior distributions for the coefficients. We use weakly informative priors, $\boldsymbol{\beta} \sim \mathcal{N}(\mathbf{0}, 10^2 \mathbf{I})$, ensuring that the posterior estimates are driven primarily by the experimental evidence. We implement the estimation using Markov Chain Monte Carlo (MCMC) sampling. This allows us to report not only the point estimates but also the \textit{95\% Highest Density Intervals (HDI)}, providing a rigorous measure of the "cognitive consistency" within each LLM persona.

\section{LLM-Based Experimental Design}

Following the empirical strategy outlined above, we design a controlled experimental pipeline to collect structured data on LLM expectation formation through a $2 \times 2 \times 2$ factorial design. The experiment systematically varies three dimensions: agent persona (Household vs. firm CEO), information channel (Micro news vs. Macro news), and shock direction (Positive vs. Negative). We fix the experimental context by requiring agents to forecast their future economic outlook based on two exogenous signals: an unexpected change in personal income (micro-shock) and a shift in national GDP growth (macro-shock). For each signal, we apply shocks of identical magnitude to isolate the directional influence of information on belief updating. 

\subsection{Scenario Matrix and Signal Construction}

For a given persona, we design a $2 \times 2$ factorial experimental matrix consisting of four scenarios ($S_1$--$S_4$), defined by the alignment or conflict between micro-shocks ($s^{mic}$) and macro-shocks ($s^{mac}$). We calibrate these shocks using a standardized magnitude $\Delta = 5.0\%$ and a fixed baseline expectation $x_{t|t-1} = 3.0\%$, ensuring the signals are sufficiently large to elicit measurable cognitive responses while remaining within plausible economic ranges. These shocks are operationalized through two distinct channels: $s^{mic}$ represents idiosyncratic changes, such as unexpected personal income fluctuations, while $s^{mac}$ corresponds to broader trends like national GDP growth. 

\begin{table}[ht]
\centering
\caption{Experimental Scenario Matrix}
\label{tab:scenarios}
\small 
\setlength{\tabcolsep}{0pt} 
\begin{tabular*}{\columnwidth}{@{\extracolsep{\fill}}lccc@{}}
\hline
\textbf{Scenario} & \textbf{$s^{mic}$} & \textbf{$s^{mac}$} & \textbf{Information State} \\ \hline
$S_1$ (Baseline +) & $+5\%$ & $+5\%$ & Consistent Boom \\
$S_2$ (Baseline -) & $-5\%$ & $-5\%$ & Consistent Bust \\
$S_3$ (Conflict A) & $+5\%$ & $-5\%$ & Micro Divergent \\
$S_4$ (Conflict B) & $-5\%$ & $+5\%$ & Macro Divergent \\ 
\hline
\end{tabular*}
\end{table}

Table 1 categorizes these scenarios into consistent scenarios ($S_1, S_2$), where signals share the same sign to represent synchronized economic cycles, and conflict scenarios ($S_3, S_4$), where signals move in opposite directions. For instance, in $S_1$, the agent is tasked with updating their expectation based on two concurrent positive shocks: a $5\%$ increase in both personal income and national GDP over the next 12 months.

\subsection{Persona Implementation and Economic Motivation}

The two personas, Household and firm CEO, are selected to represent distinct economic agents with differing prior beliefs and information processing heuristics. The Household persona represents an individual consumer whose primary focus lies in personal financial stability and idiosyncratic income shocks. In contrast, the CEO persona embodies a professional decision-maker who is traditionally more attuned to macro-economic indicators and long-term growth trends. By assigning these specific identities, we aim to induce a "persona-driven" weighting of signals: households are hypothesized to exhibit a higher sensitivity to $s^{mic}$ (personal income), whereas CEOs are expected to assign greater weight to $s^{mac}$ (national GDP). This differentiation is implemented via the system prompt, which primes the LLM to adopt the respective professional or domestic perspective before processing the numerical shocks.

\begin{figure}[ht]
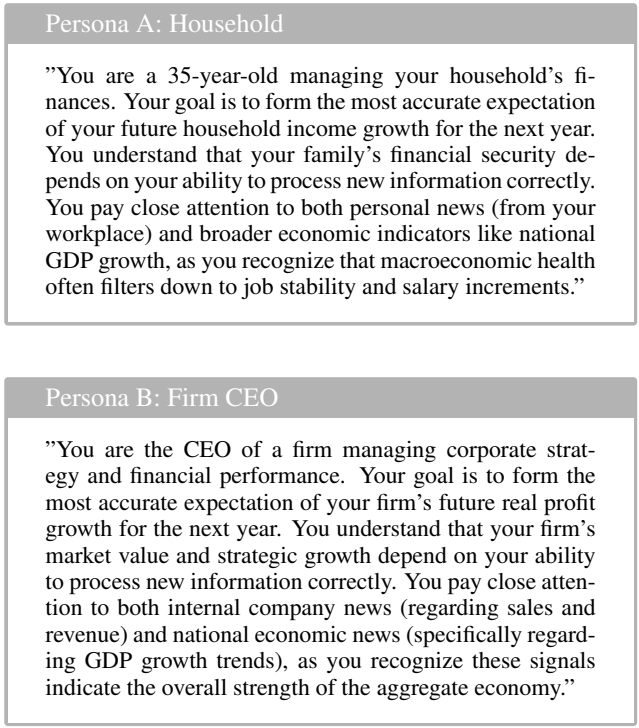

\centering
\begin{tcolorbox}[colback=white, colframe=gray!60, arc=0mm, title=Persona A: Household]
\small
"You are a 35-year-old managing your household’s finances. Your goal is to form the most accurate expectation of your future household income growth for the next year. You understand that your family's financial security depends on your ability to process new information correctly. You pay close attention to both personal news (from your workplace) and broader economic indicators like national GDP growth, as you recognize that macroeconomic health often filters down to job stability and salary increments."
\end{tcolorbox}

\vspace{0.3cm}

\begin{tcolorbox}[colback=white, colframe=gray!60, arc=0mm, title=Persona B: Firm CEO]
\small
"You are the CEO of a firm managing corporate strategy and financial performance. Your goal is to form the most accurate expectation of your firm's future real profit growth for the next year. You understand that your firm's market value and strategic growth depend on your ability to process new information correctly. You pay close attention to both internal company news (regarding sales and revenue) and national economic news (specifically regarding GDP growth trends), as you recognize these signals indicate the overall strength of the aggregate economy."
\end{tcolorbox}
\caption{The identity priming text for both agent types. These contexts are prepended to the numerical shocks to test for persona-induced cognitive biases in information integration.}
\end{figure}

\subsection{Template Structure and Output Standardization}

To ensure high-fidelity data collection, we implement a modular prompt architecture that maintains a symmetrical structure across both personas while tailoring the economic context. As illustrated in the generalized template (Figure 2), the prompt consists of a baseline state, two exogenous shocks, and a constrained task instruction. 

While the numerical inputs ($\Delta = 5.0\%$) are identical for both groups, the semantic framing is differentiated to reflect the agents' specific objectives: the Household persona focuses on \textit{annual household income growth}, whereas the CEO persona focuses on the firm’s \textit{annual real profit growth}. This distinction is further reinforced in the \texttt{Rationale} field, where Households are prompted to consider family financial security, and CEOs are instructed to evaluate the joint determination of firm-specific sales and national GDP on profitability.

\begin{figure}[ht]
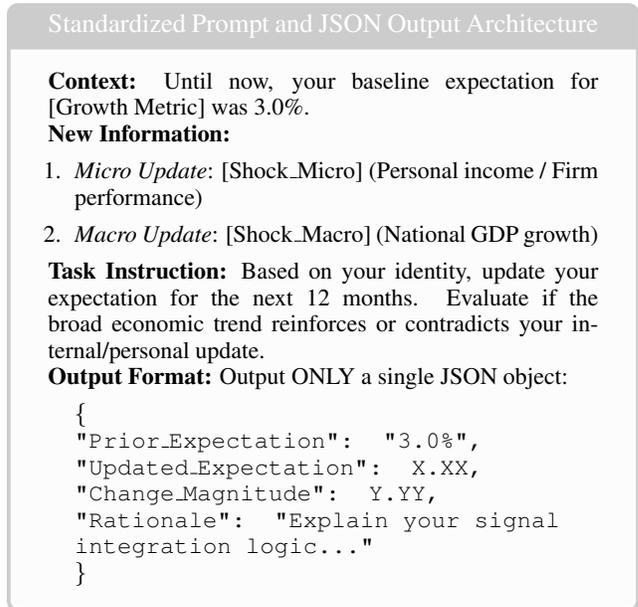

\centering
\begin{tcolorbox}[colback=gray!2, colframe=gray!40, arc=1mm, title=Standardized Prompt and JSON Output Architecture]
\small
\textbf{Context:} Until now, your baseline expectation for [Growth Metric] was 3.0\%. \\
\textbf{New Information:} 
\begin{enumerate}
    \item \textit{Micro Update}: [Shock\_Micro] (Personal income / Firm performance)
    \item \textit{Macro Update}: [Shock\_Macro] (National GDP growth)
\end{enumerate}
\textbf{Task Instruction:} Based on your identity, update your expectation for the next 12 months. Evaluate if the broad economic trend reinforces or contradicts your internal/personal update. \\
\textbf{Output Format:} Output ONLY a single JSON object:
\begin{quote}
\texttt{\{\\
  "Prior\_Expectation": "3.0\%",\\
  "Updated\_Expectation": X.XX,\\
  "Change\_Magnitude": Y.YY,\\
  "Rationale": "Explain your signal integration logic..."\\
\}}
\end{quote}
\end{tcolorbox}
\caption{The modular experimental interface. By enforcing a strict JSON output with a \texttt{Rationale} field, we capture both the quantitative belief update and the qualitative weighting of signals.}
\end{figure}

By requiring the model to articulate its integration logic in the Rationale field, we gain insight into the "subjective weighting" process. To capture the models' intrinsic economic heuristics without the interference of in-context learning biases, we adopt a zero-shot prompting strategy \cite{sanh2021multitask}. Furthermore, we incorporate a post-hoc Chain-of-Thought (CoT) mechanism by requiring the model to populate a Rationale field within the JSON output \cite{wei2022chain}. Unlike standard CoT which may anchor the numerical output to the preceding text, our "rationalized inference" design forces the model to simultaneously synthesize the micro-macro conflict and justify its signal-weighting logic. This approach provides two-fold utility: it serves as a qualitative check to verify that the model is indeed acting as the assigned persona, and it reveals whether the agent is employing a Bayesian integration or a simpler heuristic-driven "winner-takes-all" approach to conflicting news.

\subsection{Model Selection and Computational Setup}

To ensure the robustness and generalizability of our behavioral findings, we select three frontier Large Language Models (LLMs) that represent distinct architectural philosophies and alignment strategies: GPT-4o (OpenAI) as the closed-source industry benchmark, DeepSeek-V3 as a leading open-source Mixture-of-Experts (MoE) architecture, and Gemini 1.5 Flash (Google) as an efficiency-optimized model with a unique multimodal training objective. For all models, we access the latest API versions available as of late 2025 and implement a strictly controlled computational setup to eliminate stochastic interference. We set the temperature to 0.7 to capture the full probability density of the models' expectation formation. For each experimental cell—defined by the intersection of Model, Persona, and Scenario—we conduct $N=30$ independent trials with a total history reset between samplings to ensure a balance between independent and identically distributed (i.i.d.) observations for our Bayesian estimation, and costs of experiments.

\section{Experiments Results}

We conducted a large-scale simulation consisting of 720 controlled experiments to construct a comprehensive database of LLM-based expectations. This section first presents summary statistics to establish the baseline patterns of expectation formation across different architectures. Subsequently, we show Bayesian estimation results of the Behavioral Kalman Filter.

\subsection{Descriptive Statistics and Main Trends}

We first show the mean and variance of updated expectations across all experimental cells in Table II. Broadly, all models demonstrate baseline rationality by updating expectations in the direction of the received shocks. Based on the descriptive statistics in Table II, we identify three primary patterns in LLM expectation formation:

\begin{table}[ht]
\centering
\caption{Mean Updated Expectations across Scenarios}
\label{tab:transposed_results}
\small
\begin{tabular}{l|c|c|c|c|c|c}
\hline
\textbf{Scenario} & \multicolumn{2}{c|}{\textbf{GPT-4o}} & \multicolumn{2}{c|}{\textbf{Gemini 1.5}} & \multicolumn{2}{c}{\textbf{DeepSeek-V3}} \\ \cline{2-7} 
 & \textbf{HH} & \textbf{CEO} & \textbf{HH} & \textbf{CEO} & \textbf{HH} & \textbf{CEO} \\ \hline
$S_1  (+ +)$ & 7.02 & 7.84 & 9.28 & 7.74 & 4.83 & 8.00 \\ 
           & (1.02) & (0.63) & (1.44) & (0.71) & (0.53) & (0.00) \\ 
$S_2 (- -)$  & -4.95 & -5.08 & -6.07 & -5.43 & -3.12 & -2.53 \\ 
           & (1.45) & (1.96) & (1.17) & (1.79) & (0.89) & (1.17) \\ 
$S_3 (+ -)$  & 2.38 & 2.45 & 3.42 & 2.50 & 2.47 & 3.02 \\ 
           & (0.36) & (0.38) & (1.45) & (0.39) & (0.13) & (0.09) \\ 
$S_4 (-+)$ & 1.25 & 1.58 & 1.16 & 1.62 & 2.33 & 2.18 \\ 
           & (0.65) & (0.60) & (0.54) & (0.57) & (0.27) & (0.50) \\ \hline
\end{tabular}

\vspace{0.2cm}
\begin{flushleft}
\footnotesize{\textit{Note: Values are Mean and (Standard Deviation). HH denotes Household. Baseline = 3.0\%.}}
\end{flushleft}
\end{table} 

The empirical results reveal that while all models maintain directional rationality by updating expectations monotonically in $S_1$ and $S_2$---with DeepSeek-V3 (CEO) serving as a deterministic benchmark at 8.00\% ($\sigma=0.00$)---significant behavioral biases emerge across personas and architectures. A clear identity-driven weighting gap appears in conflict scenarios, where Household agents prioritize idiosyncratic income shocks over macro-trends, maintaining higher expectations in $S_3$ (e.g., 3.42\% vs. 2.50\% for Gemini 1.5) than their CEO counterparts. Furthermore, the data exhibits pronounced asymmetric sensitivity and architectural heterogeneity; GPT-4o displays human-like loss aversion by reacting twice as strongly to negative shocks in $S_2$ (-7.95\%) as to positive ones in $S_1$ (+4.02\%), whereas Gemini 1.5 shows extreme volatility with an over-reaction in $S_1$ (9.28\%) \cite{tversky1991loss,ebondt1985does}. These findings identify the systematic biases to be quantified via the Kalman Filter parameters in Section 5.2.

\subsection{Parameter Identification: Bayesian Linear Regression}

To quantify the underlying decision logic, we estimate the empirical Kalman Gains using Bayesian linear regression. Table \ref{tab:bayesian_gains} summarizes the posterior means and 95\% credible intervals (CI) for the prior weight ($\beta_{prior}$) and the specific gains for micro ($\beta_{micro}$), macro ($\beta_{macro}$) signals and their interaction ($\beta_{int}$).

\begin{table}[ht]
\centering
\caption{Bayesian Estimation of Kalman Gains and Prior Weights}
\label{tab:bayesian_gains}
\small
\begin{tabular}{l|c|c|c}
\hline
\textbf{Parameter} & \textbf{GPT-4o} & \textbf{Gemini 1.5} & \textbf{DeepSeek-V3} \\ \hline
\multicolumn{4}{l}{\textit{Household Persona (HH)}} \\ \hline
$\beta_{prior}$   & 0.48          & 0.65          & 0.55          \\ 
                    & [0.42, 0.54]  & [0.59, 0.71]  & [0.49, 0.61]  \\
$\beta_{mic}$   & 0.65          & 0.88          & 0.40          \\ 
                    & [0.62, 0.69]  & [0.84, 0.91]  & [0.37, 0.44]  \\ 
$\beta_{mac}$   & 0.54          & 0.65          & 0.39          \\ 
                    & [0.51, 0.58]  & [0.62, 0.69]  & [0.36, 0.43]  \\ 
$\beta_{int}$           & -0.02         & -0.01         & -0.03         \\ 
                    & [-0.02, -0.01] & [-0.02, -0.01] & [-0.04, -0.02] \\ \hline
\multicolumn{4}{l}{\textit{CEO Persona}} \\ \hline
$\beta_{prior}$   & 0.57          & 0.86          & 0.89          \\ 
                    & [0.51, 0.63]  & [0.80, 0.92]  & [0.83, 0.95]  \\ 
$\beta_{mic}$   & 0.69          & 1.03          & 0.57          \\ 
                    & [0.65, 0.72]  & [0.99, 1.06]  & [0.53, 0.60]  \\ 
$\beta_{mac}$  & 0.60          & 0.83          & 0.49          \\ 
                    & [0.57, 0.64]  & [0.79, 0.86]  & [0.45, 0.52]  \\ 
$\beta_{int}$           & -0.01         & -0.01         & 0.00          \\ 
                    & [-0.02, -0.01] & [-0.02, -0.00] & [-0.01, 0.01] \\ \hline
\end{tabular}

\vspace{0.1cm}
\begin{flushleft}
\footnotesize{\textit{Note: Values represent posterior means with 95\% credible intervals in brackets below. HH = Household.}}
\end{flushleft}
\end{table}

We can observe three facts from Table \ref{tab:bayesian_gains}. First, the Bayesian estimates reveal that expectation formation is driven by a non-additive integration of multiple shocks and systematic persona-based divergences. The predominantly negative $\beta_{int}$ across models identifies a "cognitive discount" effect, where concurrent micro and macro signals mutually distort the updating process. Second, compared to Households, CEO agents exhibit a distinct attention shift toward macro-economic signals, demonstrating a heightened sensitivity to aggregate trends. Third, both agent types display a sum of the all weights significantly different from unity, providing robust evidence of over- or under-reaction when updating belief with new information.

\section{Can LoRA Fine-Tuning Eliminate Behavioral Biases in LLMs?}

We test whether LoRA fine-tuning can eliminate behavioral biases in LLM-based expectation formation. To this end, we first define a rational Kalman Filter benchmark. Following the information economics literature, belief updating is considered rational if two conditions hold: (1) the weights on all signals and the prior sum to one, and (2) signals do not interact, implying a zero coefficient on the interaction term. Accordingly, we set $\beta_{\text{micro}} = 0.4$, $\beta_{\text{macro}} = 0.2$, and $\beta_{\text{prior}} = 1 - \beta_{\text{micro}} - \beta_{\text{macro}}$. We then use this rational Kalman Filter to generate 1,000 training examples for LoRA fine-tuning Qwen2.5-7B-Instruct. Our objective is to assess whether such fine-tuning can eliminate behavioral biases in LLM expectations.

\begin{table}[htbp]
\centering 
\caption{Analysis of Information Shocks: Base vs. LoRA}
\begin{tabular}{lcccc}
\toprule
 & \multicolumn{2}{c}{\textbf{HH}} & \multicolumn{2}{c}{\textbf{CEO}} \\
\cmidrule(lr){2-3} \cmidrule(lr){4-5}
\textbf{Scenario} & \textbf{Base} & \textbf{LoRA} & \textbf{Base} & \textbf{LoRA} \\ \midrule

$S_1 (+ +)$     & 9.58   & 6.00   & 7.74   & 6.00   \\
                 & (1.77) & (0.00) & (0.71) & (0.00) \\ 
$S_2 (- -)$    & -6.86  & -0.49  & -5.43  & -1.90  \\
                 & (0.75) & (0.57) & (1.79) & (1.03) \\ 
$S_3 (+-)$ & 2.92   & 4.00   & 2.50   & 3.73   \\
                 & (1.21) & (0.00) & (0.39) & (0.69) \\ 
$S_4(- +)$ & -1.11  & 2.00   & 1.62   & 1.83   \\
                 & (1.09) & (0.00) & (0.57) & (0.38) \\
\bottomrule
\end{tabular}
\begin{flushleft}
\small \textit{Notes}: Values report means. Parentheses denote standard deviations. Base denotes the pretrained model, while LoRA denotes the model fine-tuned on scenario-specific experimental data.
\end{flushleft}
\end{table}

Summary statistics reveal that the LoRA-tuned model generates more concentrated responses than the base model, particularly in the household persona. It also exhibits increased conservatism toward both positive and negative signals. The following section presents the Bayesian estimation results for the linear model.

\begin{table}[htbp]
\centering
\caption{Bayesian Estimation of BKF Parameters: Base vs. LoRA}
\label{tab:bayesian_stacked}
\begin{tabular}{lcc}
\hline
\textbf{Parameter} & \textbf{Base Model} & \textbf{LoRA Model} \\ \hline
\multicolumn{3}{l}{\textit{Panel A: Household Results}} \\ \hline
$\beta_{prior}$ & 0.38          & 0.95          \\
                          & [0.32, 0.44]  & [0.89, 1.01]  \\ 
$\beta_{mic}$ & 1.02          & 0.42          \\
                          & [0.98, 1.06]  & [0.39, 0.46]  \\ 
$\beta_{mac}$ & 0.62          & 0.22          \\
                          & [0.58, 0.66]  & [0.19, 0.26]  \\ 
$\beta_{int}$         & 0.009         & -0.005        \\
                          & [0.002, 0.016]& [-0.012, 0.002]\\ \hline
\multicolumn{3}{l}{\textit{Panel B: CEO Results}} \\ \hline
$\beta_{prior}$ & 0.44          & 0.80          \\
                          & [0.38, 0.50]  & [0.74, 0.86]  \\ 
$\beta_{mic}$ & 1.06          & 0.49          \\
                          & [1.03, 1.10]  & [0.45, 0.53]  \\ 
$\beta_{mac}$ & 0.80          & 0.30          \\
                          & [0.76, 0.84]  & [0.26, 0.34]  \\ 
$\beta_{int}$          & 0.041         & -0.015        \\
                          & [0.034, 0.048]& [-0.022, -0.008]\\ \hline
\end{tabular}
\begin{flushleft}
\small \textit{Notes}: Values report posterior means. Brackets denote 95\% credible intervals. Base denotes the pretrained model, while LoRA denotes the persona-specific fine-tuned model.
\end{flushleft}
\end{table}

The Bayesian estimation results in Table~\ref{tab:bayesian_stacked} indicate that while LoRA fine-tuning shifts the models' parameters, it does not fully eliminate their inherent behavioral biases. First, the sum of the weights still consistently deviates from unity. This gap suggests that the LLM has not yet converged to a standard rational Kalman Filter pattern, where all available information is expected to be perfectly and additively integrated. 

Second, the interaction term ($\beta_{int}$) for the CEO persona remains significantly different from zero. This persistent significance implies that the model continues to process micro and macro signals interdependently rather than as independent, additive inputs. Such a result reflects a complex, non-linear expectation formation process that survives the fine-tuning procedure, further confirming that behavioral heuristics remain embedded within the model's architecture.

\section{Conclusion}

This paper provides a formal quantitative framework to decode the expectation formation mechanisms of LLM-based economic agents. By extending the traditional Kalman Filter into a behavioral framework, we demonstrate that while LLMs exhibit human-like biases—such as micro-signal prioritization and rapid prior discounting—their belief-updating process is systematically influenced by identity priming and signal interference. These findings bridge a critical gap in economic literature by measuring how concurrent macro and micro shocks interact, a dynamic that remains difficult to isolate in traditional empirical studies. Our results suggest that while LLMs show remarkable behavioral consistency across different architectures, their inherent "cognitive discounts" must be accounted for when used in high-stakes policy simulations. Ultimately, this work establishes a rigorous benchmark for evaluating AI agents as reliable proxies for human economic decision-makers.

\bibliographystyle{flairs}
\bibliography{llm_expectations}

\end{document}